\documentclass[12pt]{myarticle}

\newtheorem{theorem}{Theorem}
\newtheorem{proposition}{Proposition}
\newtheorem{lemma}{Lemma}
\newtheorem{corollary}{Corollary}

\newtheorem{example}{Example}
\newtheorem{proof}{Proof}
\usepackage{tikz}
\usetikzlibrary{positioning}
\usepackage{subcaption}
\usepackage{amsmath}
\usepackage{breqn}
\usepackage{txfonts}
\usepackage[colorlinks]{hyperref}
\usepackage{algorithm,amsmath,algpseudocode,adjustbox,array,placeins,multirow}
\renewcommand{\d}{\displaystyle}
\newcommand{\n}{\noindent}
\usepackage{graphicx}

\begin{document}

\title{QSPR analysis of some novel neighborhood degree based topological descriptors}
\author[sm]{Sourav Mondal}
\ead{souravmath94@gmail.com}



\author[nd]{Nilanjan De}
\ead{de.nilanjan@rediffmail.com}

\author[sm]{Anita Pal}
\ead{anita.buie@gmail.com}
\address[sm]{Department of mathematics, NIT Durgapur, India.}
\address[nd]{Department of Basic Sciences and Humanities (Mathematics),\\ Calcutta Institute of Engineering and Management, Kolkata, India.}
\begin{abstract}
Topological index is a numerical value associated with chemical constitution for correlation of chemical structure with various physical properties, chemical reactivity or biological activity. In this work, some new indices based on neighbourhood degree sum of nodes are proposed. To make the computation of the novel indices convenient, an algorithm is designed. QSPR analysis of these newly introduced indices are studied here which reveals their predicting power. Some mathematical properties of these indices are also discussed here.    
\medskip

\end{abstract}


\maketitle

\section{Introduction}
The graph theory is a significant part of applied mathematics for modelling real life problems. The chemical graph theory, a fascinating branch of graph theory, provides many information on chemical compounds using an important tool called the topological index \cite{1tod00,3dev99}. Theoretical molecular descriptors alias topological indices are graph invariants that play an important role in chemistry, pharmaceutical sciences, materials science, engineering and so forth. Its role on QSPR/QSAR analysis \cite{16ran96,17ran93,dou03,dou01,bal01}, to model physical and chemical properties of molecules is also remarkable. Among several types of topological indices, vertex degree based \cite{4gut13} topological indices are most investigated and widely used. The first vertex degree based topological index is proposed in 1975 by M. Randi\'c \cite{5ran75} known as Connectivity index or Randic index. Connectivity index is defined by
\begin{eqnarray*}
R(G)= \sum\limits_{uv \in E(G)}\frac{1}{\sqrt{d_{G}(u)d_{G}(v)}}, 
\end{eqnarray*}
where $d_{G}(u)$, $d_{G}(v)$ represent the degree of nodes $u$,$v$ in the vertex set $V(G)$ of a molecular graph $G$. By molecular graph, we mean a simple connected graph considering atoms of chemical compound as vertices and the chemical bonds between them as edges. $E(G)$ is the edge set of $G$. The inverse Randic index \cite{6gut14} is given by
\begin{eqnarray*}
RR(G)=\sum\limits_{uv \in E(G)}\sqrt{d_{G}(u)d_{G}(v)}. 
\end{eqnarray*}
The Zagreb indices, introduced by Gutman and Trinajesti\'c \cite{7gut72}, are defined as follows:
\begin{eqnarray*}
 M_{1}(G) = \sum\limits_{v \in V(G)}d_{G}(v)^{2} =\sum\limits_{uv \in E(G)}[d_{G}(u) + d_{G}(v)],
\end{eqnarray*}
\begin{eqnarray*}
 M_{2}(G) &=& \sum\limits_{uv \in E(G)}[d_{G}(u)d_{G}(v)].
\end{eqnarray*}
Furtula et al. \cite{22fur15} have introduced the forgotten topological index as follows:
\begin{eqnarray*}
 F(G) = \sum\limits_{v \in V(G)}d_{G}(v)^{3} = \sum\limits_{uv \in E(G)}[d_{G}(u)^{2} + d_{G}(v)^{2}].
\end{eqnarray*}
B. Zhou and N. Trinanjsti\'c have designed the sum connectivity index \cite{8zho09} which is as follows:
\begin{eqnarray*}
SCI(G)= \sum\limits_{uv \in E(G)}\frac{1}{\sqrt{d_{G}(u)+d_{G}(v)}}. 
\end{eqnarray*}
The symmetric division degree index \cite{9vuk10} is defined as
\begin{eqnarray*}
SDD(G)= \sum\limits_{uv \in E(G)}[\frac{d_{G}(u)}{d_{G}(v)} + \frac{d_{G}(v)}{d_{G}(u)}]. 
\end{eqnarray*}
 The redefined third Zagreb index \cite{ran13} is defined by
 \begin{eqnarray*}
 ReZG_{3}(G) = \sum\limits_{uv \in E(G)}d_{G}(u)d_{G}(v)[d_{G}(u)+d_{G}(v)].
\end{eqnarray*}
For more study about degree based topological indices, readers are referred to the articles \cite{10hos17,11fur18,12de16,20hos15,sou18,21hos16}. Recently, the present authors introduced some new indices \cite{13sou18,14sou18} based on neighbourhood degree some of nodes. As a continuation, we present here some new topological indices, named as first NDe index ($ND_{1}$), second NDe index ($ND_{2}$), third NDe index ($ND_{3}$), fourth NDe index ($ND_{4}$), fifth NDe index ($ND_{5}$), and sixth NDe index ($ND_{6}$) and defined as 
\begin{dgroup*}
\begin{dmath*}
ND_{1}(G)= \sum\limits_{uv \in E(G)}\sqrt{\delta_{G}(u)\delta_{G}(v)}
\end{dmath*},
\begin{dmath*}
ND_{2}(G) =\sum\limits_{uv \in E(G)}\frac{1}{\sqrt{\delta_{G}(u)+\delta_{G}(v)}}
\end{dmath*},
\begin{dmath*}
ND_{3}(G) = \sum\limits_{uv \in E(G)}\delta_{G}(u)\delta_{G}(v)[\delta_{G}(u)+\delta_{G}(v)]
\end{dmath*},
\begin{dmath*}
ND_{4}(G) = \sum\limits_{uv \in E(G)}\frac{1}{\sqrt{\delta_{G}(u)\delta_{G}(v)}}
\end{dmath*},
\begin{dmath*}
ND_{5}(G) = \sum\limits_{uv \in E(G)}[\frac{\delta_{G}(u)}{\delta_{G}(v)} + \frac{\delta_{G}(v)}{\delta_{G}(u)}]
\end{dmath*},
\begin{dmath*}
ND_{6}(G) = \sum\limits_{uv \in E(G)}[d_{G}(u)\delta_{G}(u)+d_{G}(v)+\delta_{G}(v)]
\end{dmath*},
\end{dgroup*}
where $\delta_{G}(u)$ is the sum of degrees of all neighboring vertices of $u \in V(G)$, i.e, $\delta_{G}(u)=\sum\limits_{v \in N_{G}(u)}d_{G}(v)$, $N_{G}(u)$ being the set of adjacent vertices of u.    The goal of this article is to check the chemical applicability of the above newly designed indices and discuss about some bounds of them in terms of other topological descriptors to visualise the indices mathematically.\\
 We construct the results into two different parts. We start the first part with an algorithm for computing the indices and then some statistical regression analysis have been made to check the efficiency of the novel indices to model physical and chemical properties. Then, we would like to test their degeneracy. This part ends with a comparative study of these indices with other topological indices. The second part deals with some mathematical relation of these indices with some other well-known indices.
 \newpage
\vfill 
 \section{Computational aspects}   
In this section, we have designed an algorithm to make the computation of the novel indices convenient.

\begin{algorithm}
\caption{Computational Procedure}
\begin{algorithmic}[1]
\State \emph{Input}: Graph $G$.
\State \emph{Output}: Calculation of $\partial$ and $degree$.
\State \emph{Initialization}: $\textit{E} \gets no.~of~edges$, $\textit{V} \gets no.~of~vertex$, $\textit{conn[E][2]} \gets connection~matrix$, $\textit{deg[V][2]} \gets degree~of~each~vertex$, $\textit{$\partial$[V][2]} \gets n-bd~degree~of~each~vertex$, $\textit{ver[V]} \gets Vertex~array$, $\textit{count} \gets 0$, $\textit{adj[count]} \gets adjacent~element$, $\textit{$\partial$} \gets 0$.  
\Loop {$~i=1 ~to~ V$}
\State For each vertex from the array $ver[V]$.
\Loop {$~j=1 ~to~ E$}
\State $count$ corresponding vertex from the matrix $conn[E][2]$. 
\EndLoop
\State  $deg[V][2]=count$.
\Loop {$~k=1 ~to~ count$}
\State $adj[count]=$ store corresponding vertex.
\EndLoop
\Loop {$~k=1 ~to~ count$}
\State for each vertex from the array $adj[count]$.
\Loop {$~j=1 ~to~ E$}
\State Find the frequency of the vertex from the matrix $conn[E][2]$.
\State Store the frequency in $\partial$ for all the vertex in $adj[count]$.
\EndLoop
\State $\partial[V][2]=\partial$.
\State $\partial=0$.
\EndLoop
\State $count=0$.
\EndLoop
\State For each vertex $v \in V$.
\State Retrieve degree and n-bd degree sum from the matrix $deg[V][2]$ and $\partial[V][2]$.
\State Calculate the function $f(\delta_G(u),\delta_G(v),d_{G}(u),d_{G}(u))$. 
\end{algorithmic}
\end{algorithm}

\vfill
\clearpage
To make it simple and understandable, we have considered some variables and matrices. We have used conn $[E][2]$ matrix to store the connection details among vertices, whereas deg $[V][2]$ and $\delta[V][2]$ is the matrix to store degree of each vertex and neighborhood degree sum of vertex respectively. The novel indices can be considered as function of $\delta_{G}(u)$,$\delta_{G}(v)$,$d_{G}(u)$, and $d_{G}(v)$ i.e., f($\delta_{G}(u)$,$\delta_{G}(v)$,$d_{G}(u)$,\\
$d_{G}(v)$).
\section{Newly introduced indices in QSPR analysis}
In this section, we have studied about the newly designed topological indices to model physico-chemical properties [Acentric Factor (Acent Fac.), Entropy ($S$), enthalpy of vaporization ($HVAP$),  standard enthalpy of vaporisation ($DHVAP$), and  heat capacity at $P$ constant ($CP$)] of the octane isomers and physical properties [boiling points ($bp$), molar volumes ($mv$) at 20$^\circ$C, molar refractions ($mr$) at 20$^\circ$C., heats of vaporization (hv) at 25$^\circ$C., surface tensions (st) at 20$^\circ$C and melting points (mp)] of the 67 alkanes from n-butanes to nonanes. The experimental values of physico-chemical properties of octane isomers (Table \ref{table:1}) are taken from www.moleculardescriptors.eu. The datas related to 67 alkanes (Table \ref{table:9}) are compiled from \cite{10hos17}. Firstly, we have considered the octane isomers and then the 67 alkanes are taken into account.

\begin{table}[ht]
\caption{Experimental values of physico-chemical properties for octane isomers.}
\centering
\begin{tabular}{|c| c| c| c| c| c|}
\hline
Octanes &  Acent Fac. & S & HVAP & DHVAP & CP\\ [0.5ex]
\hline
n-octane &	0.397898 &	111.67 &	73.19 &	9.915 &	24.64\\
\hline
2-methyl heptane & 0.377916&	109.84&	70.3&	9.484&	24.8\\
\hline
3-methyl heptane &	0.371002 &	111.26 &	71.3 &	9.521 &	25.6\\
\hline
4-methyl heptane &	0.371504 &	109.32 &	70.91 &	9.483 &	25.6\\
\hline
3-ethyl hexane &	0.362472 &	109.43 &	71.7 &	9.476 &	25.74\\
\hline
2,2-dimethyl hexane &	0.339426 &	103.42 &	67.7 &	8.915 &	25.6\\
\hline
2,3-dimethyl hexane &	0.348247 &	108.02 &	70.2 &	9.272 &	26.6\\
\hline
2,4-dimethyl hexane &	0.344223 &	106.98 &	68.5 &	9.029 &	25.8\\
\hline
2,5-dimethyl hexane &	0.356830 &	105.72 &	68.6 &	9.051 &	25\\
\hline
3,3-dimethyl hexane &	0.322596 &	104.74 &	68.5 &	8.973 &	27.2\\
\hline
3,4-dimethyl hexane &	0.340345 &	106.59 &	70.2 &	9.316 &	27.4\\
\hline
2-methyl-3-ethyl pentane &	0.332433 &	106.06 &	69.7 &	9.209 &	27.4\\
\hline
3-methyl-3-ethyl pentane &	0.306899 &	101.48 &	69.3 &	9.081 &	28.9\\
\hline
2,2,3-trimethyl pentane &	0.300816 &	101.31 &	67.3 &	8.826 &	28.2\\
\hline
2,2,4-trimethyl pentane &	0.30537 &	104.09 &	64.87 &	8.402 &	25.5\\
\hline
2,3,3-trimethyl pentane &	0.293177 &	102.06 &	68.1 &	8.897 &	29\\
\hline
2,3,4-trimethyl pentane &	0.317422 &	102.39 &	68.37 &	9.014 &	27.6\\
\hline
2,2,3,3-tetramethyl butane &	0.255294 &	93.06 &	66.2 &	8.41 &	24.5\\
 [1ex]
\hline
\end{tabular}
\label{table:1}
\end{table}

\begin{table}[ht]
\caption{Topological indices of octane isomers.}
\centering
\begin{tabular}{|c| c| c| c| c| c| c|}
\hline
\textbf{Octanes} &  \textbf{$ND_{1}$} & \textbf{$ND_{2}$} & \textbf{$ND_{3}$} & \textbf{$ND_{4}$} & \textbf{$ND_{5}$} & \textbf{$ND_{6}$}\\ [1ex]
\hline
n-octane&	23.827&	2.711&	612&	2.144&	14.5&	92\\
\hline
2-methyl heptane&	25.786&	2.601&	770&	1.971&	14.517&	108\\
\hline
3-methyl heptane&	26.559&	2.5699&	892&	1.956&	15.117&	116\\
\hline
4-methyl heptane&	26.518&	2.588&	920&	1.991&	15.133&	116\\
\hline
3-ethyl hexane&	27.254&	2.551&	1056&	1.964&	15.8&	124\\
\hline
2,2-dimethyl hexane&	29.706&	2.443&	1224&	1.754&	14.6&	146\\
\hline
2,3-dimethyl hexane&	29.266&	2.444&	1212&	1.784&	15.533&	142\\
\hline
2,4-dimethyl hexane&	28.478&	2.469&	1086&	1.799&	15.183&	132\\
\hline
2,5-dimethyl hexane&	27.801&	2.495&	946&	1.802&	14.433&	124\\
\hline
3,3-dimethyl hexane&	31.1296&	2.381&	1504&	1.718&	15.933&	164\\
\hline
3,4-dimethyl hexane&	29.94&	2.404&	1332&	1.753&	16.333&	150\\
\hline
2-methyl-3-ethyl pentane&	29.902&	2.415&	1372&	1.77&	16.29&	150\\
\hline
3-methyl-3-ethyl pentane&	32.526&	2.301&	1778&	1.645&	17.364&	182\\
\hline
2,2,3-trimethyl pentane&	33.88&	2.252&	1832&	1.527&	16.107&	192\\
\hline
2,2,4-trimethyl pentane&	31.552&	2.346&	1436&	1.606&	14.752&	162\\
\hline
2,3,3-trimethyl pentane&	34.627&	2.214&	1976&	1.489&	16.681&	202\\
\hline
2,3,4-trimethyl pentane&	31.907&	2.308&	1530&	1.589&	16.057&	168\\
\hline
2,2,3,3-tetramethyl butane&	38.749&	2.076&	2534&	1.277&	15.928&	248\\
 [1ex]
\hline
\end{tabular}
\label{table:2}
\end{table}
\subsection*{Regression model for octane isomers:}
We have tested the following linear regression models
\begin{eqnarray}
P = m(TI)+c,
\end{eqnarray}
  where P is the physical property and TI is the topological index. Using the above formula, we have the following linear regression models for different neighbourhood degree based topological indices.\\
1. $ND_{1}$ index:\\
\begin{align*}
S &=141.1521-[ND_{1} (G)]1.1926\\
Acent~Fac.&=0.627 - [ND_{1} (G)]0.0097\\
DHVAP&=11.8017 - [ND_1 (G)]0.0893
\end{align*}

2. $ND_{2}$ index:\\
\begin{align*}
S&=39.6776+[ND_2 (G)]27.1579\\
Acent ~Fac.&=-0.2058 + [ND_2 (G)]0.2238\\
DHVAP&=1.1069 + [ND_2 (G)]2.0737
\end{align*}

3. $ND_{3}$ index:\\
\begin{align*}
S&=117.2259-[ND_3 (G)]0.0088\\
Acent~ Fac.&= 0.4322 - [ND_3 (G)]7.2 \times 10^{-5}\\
DHVAP&=9.9568 - [ND_3 (G)]0.0006
\end{align*}

4. $ND_{4}$ index:\\
\begin{align*}
S&=69.8183+[ND_4 (G)]20.3149\\
Acent~ Fac.&= 0.04656+ [ND_4 (G)]0.1651\\
DHVAP&= 6.2868 + [ND_4 (G)]1.6206\\
HVAP&= 55.1172 + [ND_4 (G)]8.0164
\end{align*}

5. $ND_{5}$ index:\\
\begin{align*}
S&=144.7836-[ND_5 (G)]2.5286\\
Acent ~Fac.&= 0.7245 - [ND_5 (G)]0.0249\\
DHVAP&= 10.6958 - [ND_5 (G)]0.1008\\
CP&=3.8987+[ND_5 (G)]1.4447
\end{align*}

6. $ND_{6}$ index:\\
\begin{align*}
S&=122.3482-[ND_6 (G)]0.1122\\
Acent~ Fac.&= 0.4730 -[ND_6 (G)]0.0009\\
DHVAP&= 10.3438 - [ND_6 (G)]0.0081
\end{align*}

Now we describe above linear models in the following tableau. Here c, m, r, SE, F, SF stands for intercept, slope, correlation coefficient, standard error, F-test, and significance F respectively. Correlation coefficient tells how strong the linear relationship is. The standard error of the regression is the precision that the regression coefficient is measured. To check whether the results are reliable, Significance F can be useful. If this value is less than 0.05, then the model is statistically significant. If significance F is greater than 0.05, it is probably better to stop using that set of independent variable.

\begin{table}[ht]
\caption{Statical parameters for the linear QSPR model for $ND_{1}(G)$.}
\centering
\begin{tabular}{|c| c| c| c| c| c| c|}
\hline

Physical Properties&	c&	m&	r&	SE&	F&	SF\\
\hline
S&	141.1521&	-1.1926&	-0.9537&	1.4010&	160.7549&	9.2E-10\\
\hline

Acent Fac.&	0.627&	-0.0097&	-0.9904&	0.0050&	824.2198&	3.42E-15\\
\hline

DHVAP&	11.8017&	-0.0893&	-0.8414&	0.2135&	38.7783&	1.21E-05\\
\hline
\end{tabular}
\label{table:3}
\end{table}

\begin{table}[ht]
\caption{Statical parameters for the linear QSPR model for $ND_2 (G)$.}
\centering
\begin{tabular}{|c| c| c| c| c| c| c|}
\hline

Physical Properties&	c&	m&	r&	SE&	F&	SF\\
\hline
S&	39.6776&	27.1579&	0.9419&	1.5629&	126.0258&	5.37E-09\\
\hline
Acent Fac.&	-0.2058&	0.2238&	0.9890&	0.0054&	717.1224&	1.02E-14\\
\hline
DHVAP&	1.1069&	2.0737&	0.8477&	0.2096&	40.8614&	8.92E-06\\

\hline
\end{tabular}
\label{table:4}
\end{table}

\begin{table}[ht]
\caption{Statical parameters for the linear QSPR model for $ND_3 (G)$.}
\centering
\begin{tabular}{|c| c| c| c| c| c| c|}
\hline

Physical Properties&	c&	m&	r&	SE&	F&	SF\\
\hline
S&	117.2259&	-0.0088&	-0.9387&	1.6052&	118.6526&	8.25E-09\\
\hline
Acent Fac.&	0.4322&	-7.2E-05&	-0.9765&	0.0079&	328.2359&	4.37E-12\\
\hline
DHVAP&	9.9568&	-0.0006&	-0.7778&	0.2483&	24.5074&	0.000145\\

\hline
\end{tabular}
\label{table:5}
\end{table}
\newpage
\vfill

\begin{table}[ht]
\caption{Statical parameters for the linear QSPR model for $ND_4 (G)$.}
\centering
\begin{tabular}{|c| c| c| c| c| c| c|}
\hline

Physical Properties&	c&	m&	r&	SE&	F&	SF\\
\hline
S&	69.8183&	20.3149&	0.9491&	1.4667&	145.2802&	1.93E-09\\
\hline
Acent Fac.&	0.0465&	0.1651&	0.9830&	0.0067&	458.9889&	3.31E-13\\
\hline
DHVAP&	6.2868&	1.6206&	0.8923&	0.1784&	62.5202&	6.45E-07\\
\hline
HVAP&	55.1172&	8.0164&	0.8350&	1.1493&	36.8449&	1.62E-05\\
\hline
\end{tabular}
\label{table:6}
\end{table}

\begin{table}[ht]
\caption{Statical parameters for the linear QSPR model for $ND_5 (G)$.}
\centering
\begin{tabular}{|c| c| c| c| c| c| c|}
\hline

Physical Properties&	c&	m&	r&	SE&	F&	SF\\
\hline
CP&	3.8987&	1.4447&	0.8478&	0.7856&	40.9017&	8.87E-06\\
\hline
S&	144.7836&	-2.5286&	-0.4721&	4.1049&	4.5895&	0.047895\\
\hline
Acent Fac.&	0.7245&	-0.0249&	-0.2218&	0.0294&	8.7253&	0.009335\\
\hline
DHVAP&	10.6958&	-0.1008&	-0.5940&	0.3853&	0.8280&	0.376344\\

\hline
\end{tabular}
\label{table:7}
\end{table}

\begin{table}[ht]
\caption{Statical parameters for the linear QSPR model for $ND_6 (G)$.}
\centering
\begin{tabular}{|c| c| c| c| c| c| c|}
\hline

Physical Properties&	c&	m&	r&	SE&	F&	SF\\
\hline
S&	122.3482&	-0.1122&	-0.9508&	1.4420&	150.8369&	1.47E-09\\
\hline
Acent Fac.&	0.4730&	-0.0009&	-0.9821&	0.0069&	434.0329&	5.09E-13\\
\hline
DHVAP&	10.3438&	-0.0081&	-0.8056&	0.2341&	29.5844&	5.46E-05\\
\hline
\end{tabular}
\label{table:8}
\end{table}
Now we depict the above correlations in the following figures.

\FloatBarrier
\begin{figure}[ht]
\includegraphics[height= 4cm,width=15cm]{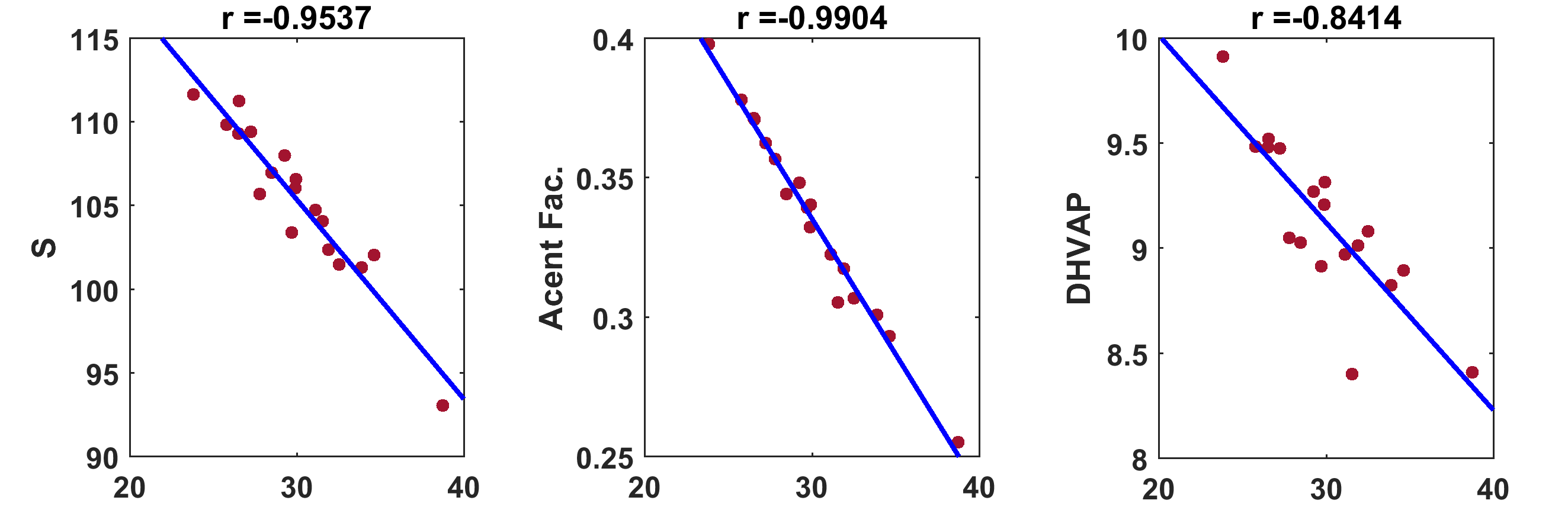}
\centering
\caption{Correlation of $ND_{1}$ index with S, Acent Fac., and DHVAP for octane isomers.}
\label{fig1}
\end{figure}

\begin{figure}[ht]
\includegraphics[height= 4cm,width=15cm]{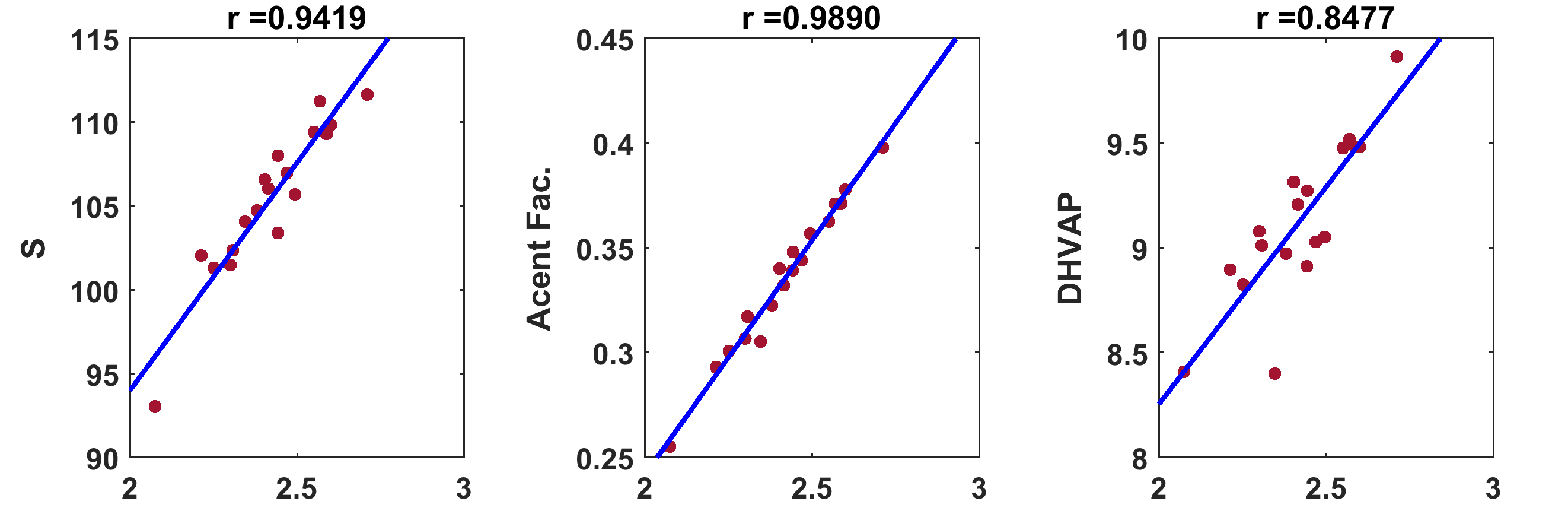}
\centering
\caption{Correlation of $ND_{2}$ index with S, Acent Fac., and DHVAP for octane isomers.}
\label{fig2}
\end{figure}

\begin{figure}[ht]
\includegraphics[height= 4cm,width=15cm]{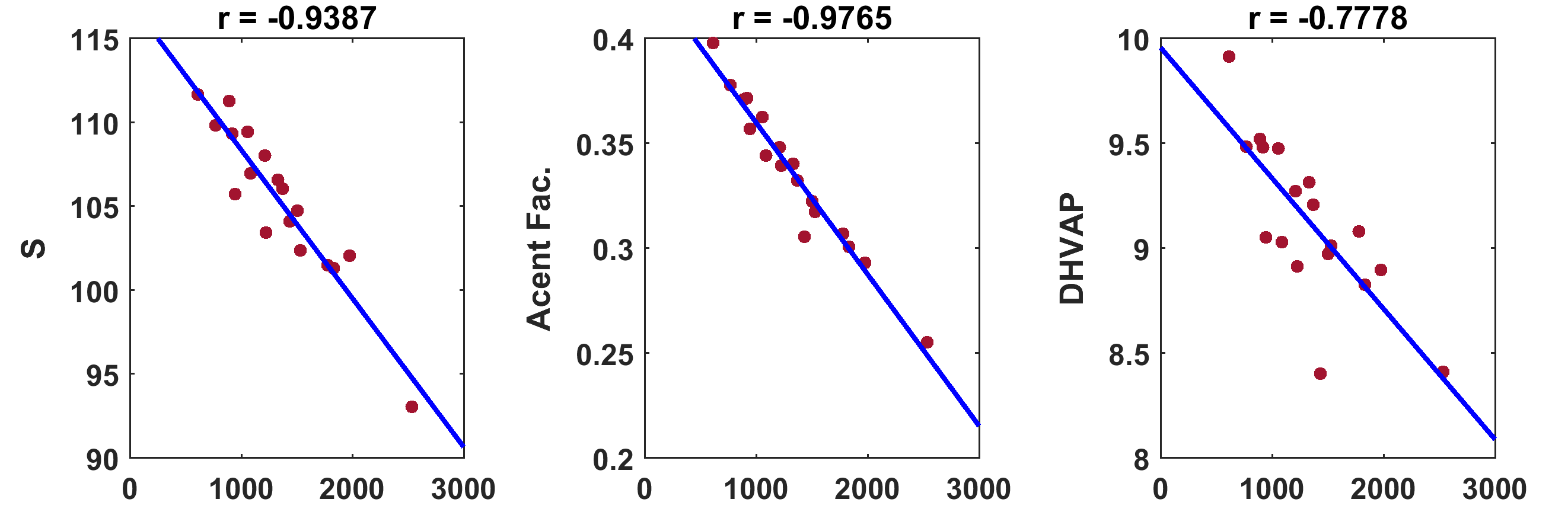}
\centering
\caption{Correlation of $ND_{3}$ index with S, Acent Fac., and DHVAP for octane isomers.}
\label{fig3}
\end{figure}

\begin{figure}[ht]
\includegraphics[height= 5.5cm,width=14cm]{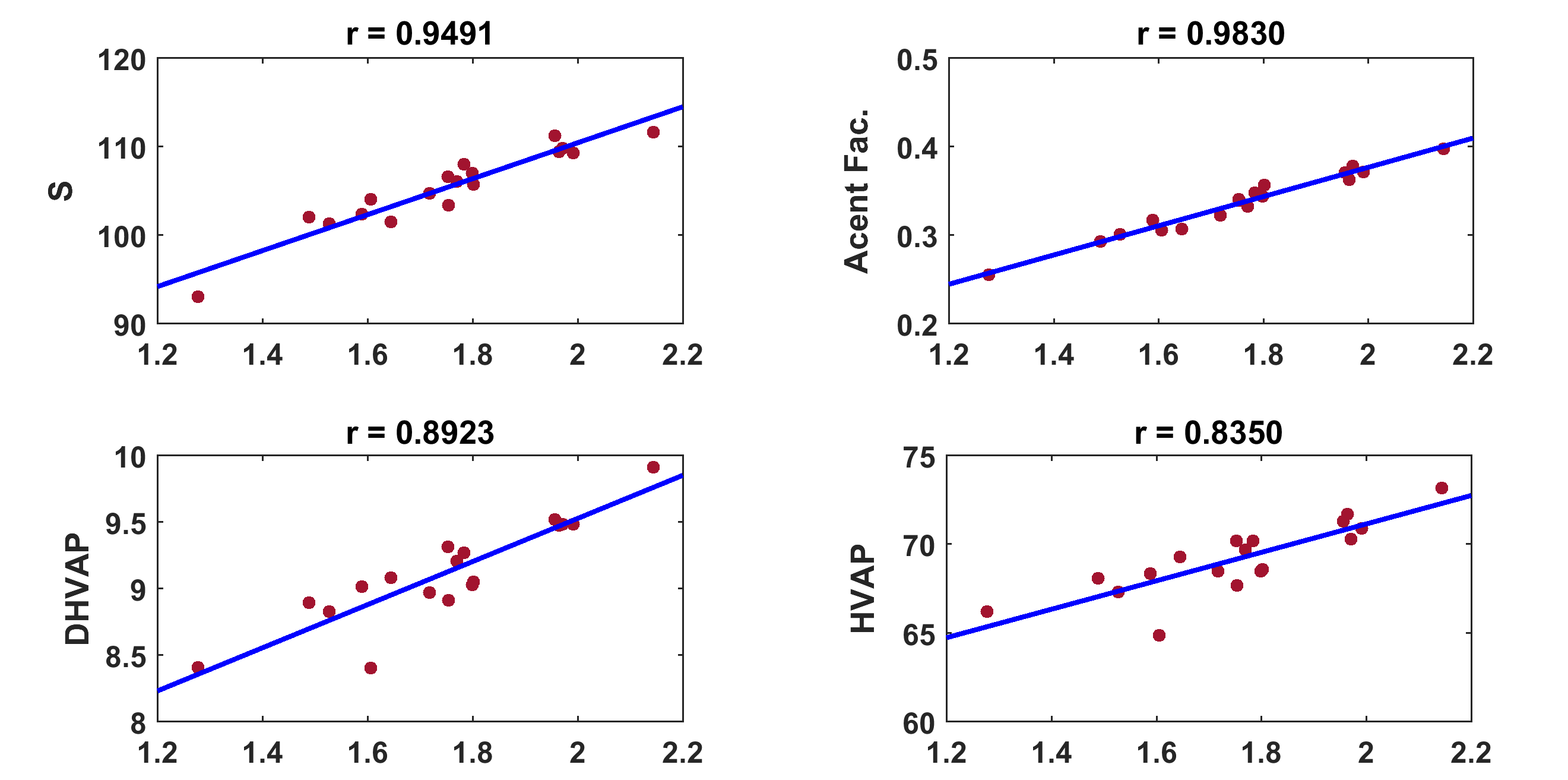}
\centering
\caption{Correlation of $ND_{4}$ index with S, Acent Fac., and DHVAP for octane isomers.}
\label{fig4}
\end{figure}
\begin{figure}[ht]
\includegraphics[height=5.5cm,width=14cm]{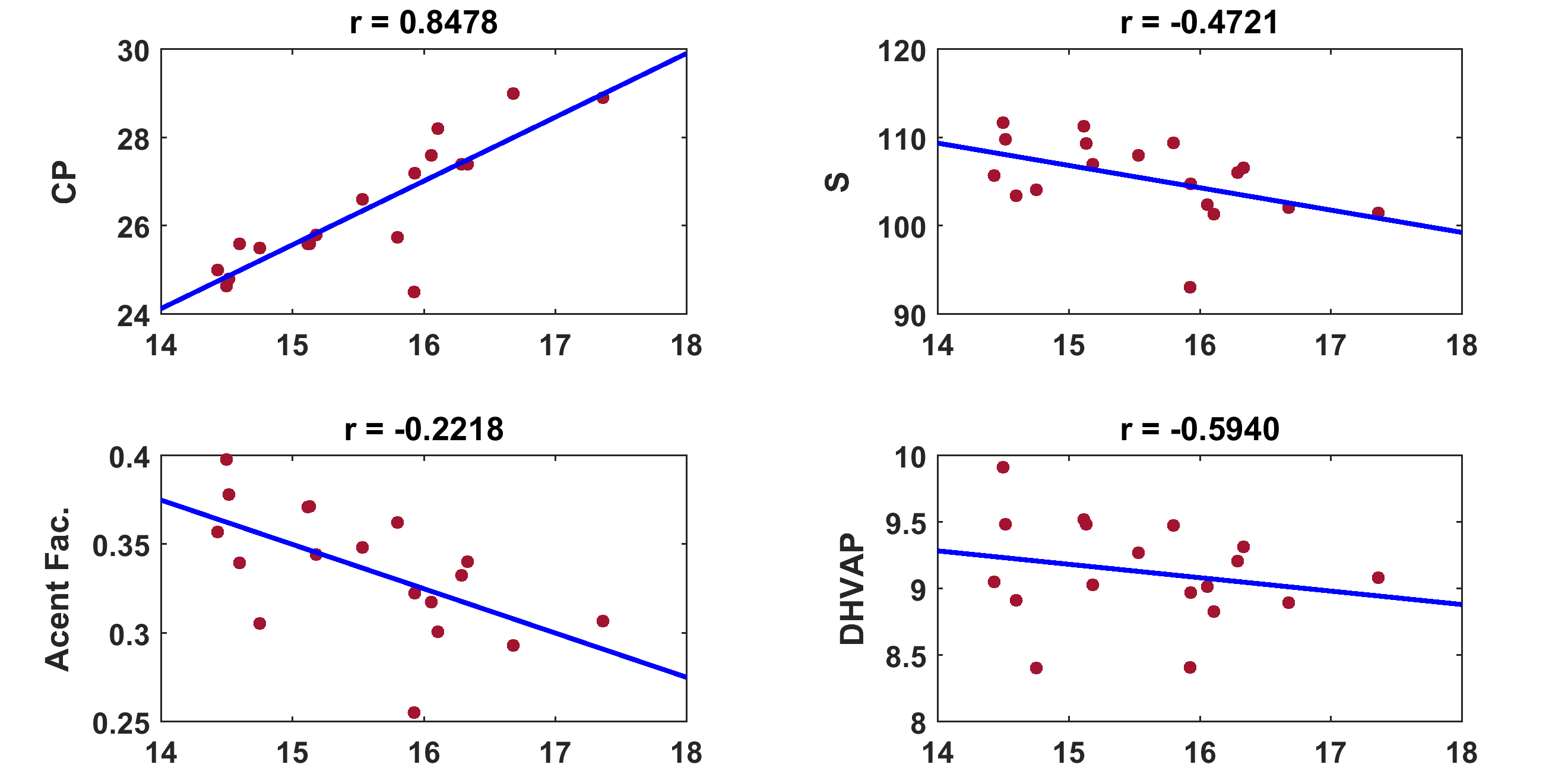}
\centering
\caption{Correlation of $ND_{5}$ index with S, Acent Fac., and DHVAP for octane isomers.}
\label{fig5}
\end{figure}

\begin{figure}[ht]
\includegraphics[height= 3.3cm,width=15cm]{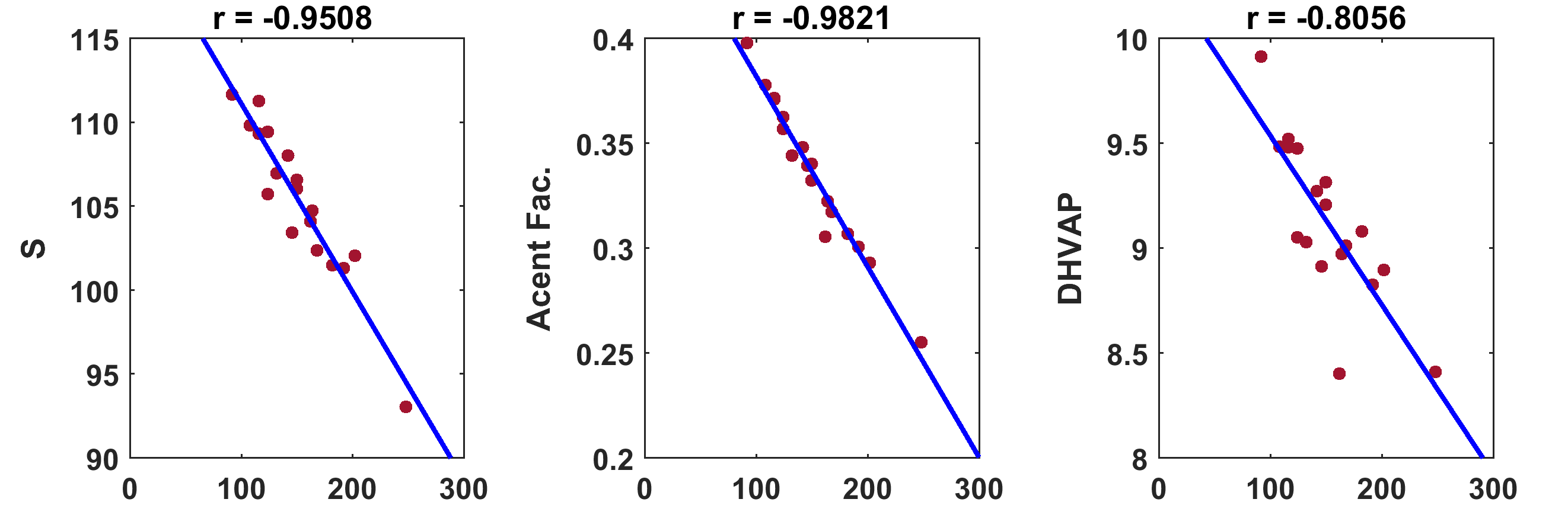}
\centering
\caption{Correlation of $ND_{6}$ index with S, Acent Fac., and DHVAP for octane isomers.}
\label{fig6}
\end{figure}
\vfill
\clearpage

\FloatBarrier
\begin{table}[ht]
\caption{Experimental values of physical properties for 67 alkanes.}
\centering
\begin{adjustbox}{width=1.2\textwidth}
\begin{tabular}{|c| c| c| c| c| c| c| c| c|}
\hline

Alkanes&	\textbf{$bp(^\circ$$C$)}&	\textbf{$mv(cm^{3})$}	&\textbf{$mr(cm^{3}$)}&	\textbf{$hv(kJ)$}&	\textbf{$ct$($^\circ$$C$)}&	\textbf{$cp(atm)$}&	\textbf{$st(dyne/cm)$}&	\textbf{$mp$($^\circ$$C$)}\\
\hline
Butane&	-0.05&	& & &			152.01&	37.47&	&	-138.35\\
\hline
2-methyl propane&	-11.73& &&&				134.98&	36&		&-159.6\\
\hline
Pentane&	36.074&	115.205&	25.2656&	26.42&	196.62&	33.31&	16&	-129.72\\
\hline
2-methyl butane&	27.852&	116.426&	25.2923&	24.59&	187.7&	32.9&	15&	-159.9\\
\hline
2,2-dimethyl propane&	9.503&	112.074&	25.7243&	21.78&	160.6&	31.57&	& 	-16.55\\
\hline
Hexane&	68.74&	130.688&	29.9066&	31.55&	234.7&	29.92&	18.42&	-95.35\\
\hline
2-methyl pentane&	60.271&	131.933&	29.945&	29.86&	224.9&	29.95&	17.38&	-153.67\\
\hline
3-methyl pentane&	63.282&	129.717&	29.8016&	30.27&	231.2&	30.83&	18.12&	-118\\
\hline
2,2-methyl butane&	4.741&	132.744&	29.9347&	27.69&	216.2&	30.67&	16.3&	-99.87\\
\hline
2,3-dimethyl butane&	57.988&	130.24&	29.8104&	29.12&	227.1&	30.99&	17.37&	-128.54\\
\hline
Heptane&	98.427&	146.54&	34.5504&	36.55&	267.55&	27.01&	20.26&	-90.61\\
\hline
2-methyl hexane&	90.052&	147.656&	34.5908&	34.8&	257.9&	27.2&	19.29&	-118.28\\
\hline
3-methyl hexane&	91.85&	145.821&	34.4597&	35.08&	262.4&	28.1&	19.79&	-119.4\\
\hline
3-ethyl pentane&	93.475&	143.517&	34.2827&	35.22&	267.6&	28.6&	20.44&	-118.6\\
\hline
2,2-dimethyl pentane&	79.197&	148.695&	34.6166&	32.43&	247.7&	28.4&	18.02&	-123.81\\
\hline
2,3-dimethyl pentane&	89.784&	144.153&	34.3237&	34.24&	264.6&	29.2&	19.96&	-119.1\\
\hline
2,4-dimethyl pentane&	80.5&	148.949&	34.6192&	32.88&	247.1&	27.4&	18.15&	-119.24\\
\hline
3,3-dimethyl pentane&	86.064&	144.53&	34.3323&	33.02&	263&	30&	19.59&	-134.46\\
\hline
Octane&	125.665&	162.592&	39.1922&	41.48&	296.2&	24.64&	21.76&	-56.79\\
\hline
2-methyl heptane&	117.647&	163.663&	39.2316&	39.68&	288&	24.8&	20.6&	-109.04\\
\hline
3-methyl heptane&	118.925&	161.832&	39.1001&	39.83&	292&	25.6&	21.17&	-120.5\\
\hline
3-methyl heptane&	117.709&	162.105&	39.1174&	39.67&	290&	25.6&	21&	-120.95\\
\hline
3-ethyl hexane&	118.53&	160.07&	38.94&	39.4&	292&	25.74&	21.51&  \\
\hline	
2,2-dimethyl hexane&	10.84&	164.28&	39.25&	37.29&	279&	25.6&	19.6&	-121.18\\
\hline
2,3-dimethyl hexane&	115.607&	160.39&	38.98&	38.79&	293&	26.6&	20.99&  \\	
\hline
2,4-dimethyl hexane&	109.42&	163.09&	39.13&	37.76&	282&	25.8&	20.05&	-137.5\\
\hline
2,5-dimethyl hexane&	109.1&	164.69&	39.25&	37.86&	279&	25&	19.73&	-91.2\\
\hline
3,3-dimethyl hexane&	111.96&	160.87&	39&	37.93&	290.84&	27.2&	20.63&	-126.1\\
\hline
3,4-dimethyl hexane&	117.72&	158.81&	38.84&	39.02&	298&	27.4&	21.64& \\	
\hline
3-ethyl-2-methyl pentane&	115.65&	158.79&	38.83&	38.52&	295&	27.4&	21.52&	-114.96\\
\hline
3-ethyl-3-methyl pentane&	118.25&	157.02&	38.71&	37.99&	305&	28&	21.99&	-90.87\\
\hline
2,2,3-trimethyl pentane&	109.84&	159.52&	38.92&	36.91&	294&	28.2&	20.67&	-112.27\\
\hline
2,2,4-trimethyl pentane&	99.23&	165.08&	39.26&	35.13&	271.15&	25.5&	18.77&	-107.38\\
\hline
2,3,3-trimethyl pentane&	114.76&	157.29&	38.76&	37.22&	303&	29&	21.56&	-100.7\\
\hline
2,3,4-trimethyl pentane&	113.46&	158.85&	38.86&	37.61&	295&	27.6&	21.14&	-109.21\\
\hline
Nonane&	150.76&	178.71&	43.84&	46.44&	322&	22.74&	22.92&	-53.52\\
\hline
2-methyl octane&	143.26&	179.77&	43.87&	44.65&	315&	23.6&	21.88&	-80.4\\
\hline
3-methyl octane&	144.18&	177.5&	43.72&	44.75&	318&	23.7&	22.34&	-107.64\\
\hline

\end{tabular}
\end{adjustbox}
\label{table:9}
\end{table}

\begin{table*}[ht]
\centering
\begin{adjustbox}{width=1.2\textwidth}
\begin{tabular}{|c| c| c| c| c| c| c| c| c|}
\hline

4-methyl octane&	142.48&	178.15&	43.76&	44.75&	318.3&	23.06&	22.34&	-113.2\\
\hline
3-ethyl heptane&	143&	176.41&	43.64&	44.81&	318&	23.98&	22.81&	-114.9\\
\hline
4-ethyl heptane&	141.2&	175.68&	43.69&	44.81&	318.3&	23.98&	22.81& \\
\hline	
2,2-dimethyl heptane&	132.69&	180.5&	43.91&	42.28&	302&	22.8&	20.8&	-113\\
\hline
2,3-dimethyl heptane&	140.5&	176.65&	43.63&	43.79&	315&	23.79&	22.34&	-116\\
\hline
2,4-dimethyl heptane&	133.5&	179.12&	43.73&	42.87&	306&	22.7&	23.3&	\\
\hline
2,5-dimethyl heptane&	136&	179.37&	43.84&	43.87&	307.8&	22.7&	21.3& \\	
\hline
2,6-dimethyl heptane&	135.21&	180.91&	43.92&	42.82&	306&	23.7&	20.83&	-102.9\\
\hline
3,3-dimethyl heptane&	137.3&	176.897&	43.687&	42.66&	314&	24.19&	22.01& \\	
\hline
3,4-dimethyl heptane&	140.6&	175.349&	43.5473&	43.84&	322.7&	24.77&	22.8& \\
\hline	
3,5-dimethyl heptane&	136&	177.386&	43.6379&	42.98&	312.3&	23.59&	21.77& \\
\hline	
4,4-dimethyl heptane&	135.2&	176.897&	43.6022&	42.66&	317.8&	24.18&	22.01& \\	
\hline
3-ethyl-2-methyl hexane&	138&	175.445&	43.655&	43.84&	322.7&	24.77&	22.8& \\	
\hline
4-ethyl-2-methyl hexane&	133.8&	177.386&	43.6472&	42.98&	330.3&	25.56&	21.77& \\	
\hline
3-ethyl-3-methyl hexane&	140.6&	173.077&	43.268&	44.04&	327.2&	25.66&	23.22& \\	
\hline
2,2,4-trimethyl hexane&	126.54&	179.22&	43.7638&	40.57&	301&	23.39&	20.51&	-120\\
\hline
2,2,5-trimethyl hexane&	124.084&	181.346&	43.9356&	40.17&	296.6&	22.41&	20.04&	-105.78\\
\hline
2,3,3-trimethyl hexane&	137.68&	173.78&	43.4347&	42.23&	326.1&	25.56&	22.41&	-116.8\\
\hline
2,3,4-trimethyl hexane&	139&	173.498&	43.4917&	42.93&	324.2&	25.46&	22.8& \\
\hline	
2,3,5-trimethyl hexane&	131.34&	177.656&	43.6474&	41.42&	309.4&	23.49&	21.27&	-127.8\\
\hline
3,3,4-trimethyl hexane&	140.46&	172.055&	43.3407&	42.28&	330.6&	26.45&	23.27&	-101.2\\
\hline
3,3-diethyl pentane&	146.168&	170.185&	43.1134&	43.36&	342.8&	26.94&	23.75&	-33.11\\
\hline
2,2-dimethyl-3-ethyl pentane&	133.83&	174.537&	43.4571&	42.02&	322.6&	25.96&	22.38&	-99.2\\
\hline
2,3-dimethyl-3-ethyl pentane&	142&	170.093&	42.9542&	42.55&	338.6&	26.94&	23.87& \\
\hline	
2,4-dimethyl-3-ethyl pentane&	136.73&	173.804&	43.4037&	42.93&	324.2&	25.46&	22.8&	-122.2\\
\hline
2,2,3,3-tetramethyl pentane&	140.274&	169.495&	43.2147&	41&	334.5&	27.04&	23.38&	-99\\
\hline
2,2,3,4-tetramethyl pentane&	133.016&	173.557&	43.4359&	41&	319.6&	25.66&	21.8&	-121.09\\
\hline
2,2,4,4-tetramethyl pentane&	122.284&	178.256&	43.8747&	38.1&	301.6&	24.58&	20.37&	-66.54\\
\hline
2,3,3,4-tetramethyl pentane&	141.551&	169.928&	43.2016&	41.75&	334.5&	26.85&	23.31&	-102.12\\
\hline

\hline
\end{tabular}
\end{adjustbox}
\end{table*}
\FloatBarrier

\FloatBarrier
\begin{table}[ht]
\caption{Topological indices for 67 alkanes.}
\centering
\begin{adjustbox}{width=1\textwidth}
\begin{tabular}{|c| c| c| c| c| c| c|}
\hline

Alkanes&	$ND_{1}$&	$ND_{2}$&	$ND_{3}$&	$ND_{4}$&	$ND_{5}$&	$ND_{6}$\\
\hline
Butane&	7.899&	1.303&	114&	1.149&	6.333&	28\\
\hline
2-methyl propane&	9&	1.225&	16&2	1&	6&	36\\
\hline
Pentane&	11.827&	1.65&	228&	1.394&	8.5&	44\\
\hline
2-methyl butane&	13.757&	1.518&	344&	1.181&	8.667&	60\\
\hline
2,2-dimethyl propane&	16&	1.414&	512&	1&	8&	80\\
\hline
Hexane&	15.827&	2.004&	356&	1.644&	10.5&	60\\
\hline
2-methyl pentane&	17.723&	1.89&	498&	1.467&	10.65&	76\\
\hline
3-methyl pentane&	18.474&	1.837&	576&	1.412&	11.367&	84\\
\hline
2,2-methyl butane&	21.579&	1.694&	860&	1.187&	11.05&	114\\
\hline
2,3-dimethyl butane&	20.492&	1.73&	730&	1.233&	11.067&	102\\
\hline
Heptane&	19.827&	2.357&	484&	1.894&	12.5&	76\\
\hline
2-methyl hexane&	21.786&	2.248&	642&	1.721&	12.517&	92\\
\hline
3-methyl hexane&	22.496&	2.212&	748&	1.702&	13.25&	100\\
\hline
3-ethyl pentane&	23.182&	2.173&	864&	1.673&	14&	108\\
\hline
2,2-dimethyl pentane&	25.586&	2.082&	1062&	1.497&	12.85&	130\\
\hline
2,3-dimethyl pentane&	25.193&	2.066&	1020&	1.492&	13.733&	126\\
\hline
2,4-dimethyl pentane&	23.654&	2.144&	816&	1.563&	12.667&	108\\
\hline
3,3-dimethyl pentane&	31.129&	2.381&	1504&	1.718&	15.933&	164\\
\hline
Octane&	23.827&	2.711&	612&	2.144&	14.5&	92\\
\hline
2-methyl heptane&	25.786&	2.601&	770&	1.971&	14.517&	108\\
\hline
3-methyl heptane&	26.559&	2.569&	892&	1.956&	15.117&	116\\
\hline
3-methyl heptane&	26.518&	2.588&	920&	1.991&	15.133&	116\\
\hline
3-ethyl hexane&	24.451&	2.658&	682&	2.097&	15.483&	106\\
\hline
2,2-dimethyl hexane&	29.706&	2.443&	1224&	1.754&	14.6&	146\\
\hline
2,3-dimethyl hexane&	29.266&	2.444&	1212&	1.784&	15.533&	142\\
\hline
2,4-dimethyl hexane&	28.478&	2.469&	1086&	1.799&	15.183&	132\\
\hline
2,5-dimethyl hexane&	27.801&	2.495&	946&	1.802&	14.433&	124\\
\hline
3,3-dimethyl hexane&	31.129&	2.381&	1504&	1.718&	15.933&	164\\
\hline
3,4-dimethyl hexane&	29.94&	2.404&	1332&	1.753&	16.333&	150\\
\hline
3-ethyl-2-methyl pentane&	29.902&	2.415&	1372&	1.77&	16.29&	150\\
\hline
3-ethyl-3-methyl pentane&	32.526&	2.301&	1778&	1.645&	17.364&	182\\
\hline
2,2,3-trimethyl pentane&	33.88&	2.252&	1832&	1.527&	16.107&	192\\
\hline

\end{tabular}
\end{adjustbox}
\label{table:10}
\end{table}

\begin{table*}[ht]
\centering
\begin{adjustbox}{width=1\textwidth}
\begin{tabular}{|c| c| c| c| c| c| c|}
\hline

2,2,4-trimethyl pentane&	31.552&	2.346&	1436&	1.606&	14.752&	162\\
\hline
2,3,3-trimethyl pentane&	34.627&	2.214&	1976&	1.489&	16.681&	202\\
\hline
2,3,4-trimethyl pentane&	31.907&	2.308&	1530&	1.589&	16.057&	168\\
\hline
Nonane&	27.827&	3.064&	740&	2.394&	16.5&	108\\
\hline
2-methyl octane&	29.786&	2.955&	898&	2.221&	16.517&	124\\
\hline
3-methyl octane&	30.559&	2.923&	1020&	2.206&	17.117&	132\\
\hline
4-methyl octane&	29.116&	3.004&	888&	2.323&	16.917&	123\\
\hline
3-ethyl heptane&	31.318&	2.909&	1200&	2.218&	17.667&	140\\
\hline
4-ethyl heptane&	30.292&	2.985&	1152&	2.342&	17.6&	136\\
\hline
2,2-dimethyl heptane&	33.706&	2.796&	1352&	2.004&	16.6&	162\\
\hline
2,3-dimethyl heptane&	33.329&	2.802&	1356&	2.038&	17.4&	158\\
\hline
2,4-dimethyl heptane&	32.499&	2.844&	1258&	2.089&	17.067&	148\\
\hline
2,5-dimethyl heptane&	32.574&	2.817&	1196&	2.036&	17.033&	148\\
\hline
2,6-dimethyl heptane&	31.745&	2.845&	1056&	2.049&	16.533&	140\\
\hline
3,3-dimethyl heptane&	35.25&	2.742&	1666&	1.975&	17.683&	180\\
\hline
3,4-dimethyl heptane&	34.012&	2.782&	1524&	2.045&	18.133&	166\\
\hline
3,5-dimethyl heptane&	34.179&	2.768&	1536&	2.009&	17.862&	160\\
\hline
4,4-dimethyl heptane&	35.182&	2.771&	1728&	2.029&	17.667&	180\\
\hline
3-ethyl-2-methyl hexane&	34.02&	2.795&	1586&	2.063&	18.017&	166\\
\hline
4-ethyl-2-methyl hexane&	33.282&	2.809&	1416&	2.063&	17.667&	156\\
\hline
3-ethyl-3-methyl hexane&	36.621&	2.693&	2026&	1.958&	19.04&	198\\
\hline
2,2,4-trimethyl hexane&	36.422&	2.672&	1728&	1.837&	17.195&	186\\
\hline
2,2,5-trimethyl hexane&	35.771&	2.692&	1548&	1.837&	16.433&	178\\
\hline
2,3,3-trimethyl hexane&	38.722&	2.605&	2224&	1.802&	18.357&	218\\
\hline
2,3,4-trimethyl hexane&	36.695&	2.647&	1866&	1.851&	18.6&	192\\
\hline
2,3,5-trimethyl hexane&	35.293&	2.703&	1572&	1.883&	17.4&	174\\
\hline
3,3,4-trimethyl hexane&	39.364&	2.563&	2358&	1.767&	19.24&	226\\
\hline
3,3-diethyl pentane&	37.947&	2.621&	2360&	1.897&	20.5&	216\\
\hline
2,2-dimethyl-3-ethyl pentane&	38.596&	2.609&	2256&	1.817&	18.583&	216\\
\hline
2,3-dimethyl-3-ethyl pentane&	40.044&	2.533&	2560&	1.741&	19.833&	236\\
\hline
2,4-dimethyl-3-ethyl pentane&	36.626&	2.666&	1952&	1.879&	18.517&	192\\
\hline
2,2,3,3-tetramethyl pentane&	44.158&	2.395&	3122&	1.528&	19.107&	282\\
\hline
2,2,3,4-tetramethyl pentane&	40.595&	2.502&	2416&	1.635&	18.383&	234\\
\hline
2,2,4,4-tetramethyl pentane&	39.482&	2.555&	2120&	1.658&	16.75&	216\\
\hline
2,3,3,4-tetramethyl pentane&	42.141&	2.445&	2760&	1.585&	19.167&	238\\
\hline
\end{tabular}
\end{adjustbox}
\end{table*}
\FloatBarrier
\subsection*{Regression model for 67 alkanes:}
1. $ND_{1}$ index:\\
\begin{align*}
bp&=-8.8069+[ND_1 (G)]4.0114\\
ct&=135.4475+[ND_1 (G)]5.1317\\
cp&=34.6560-[ND_1 (G)]0.2721\\
mv&=100.8619+[ND_1 (G)]2.0398\\
mr&=20.1480+[ND_1 (G)]0.6398\\
hv&=21.8703+[ND_1 (G)]0.5616\\
st&=14.3557+[ND_1 (G)]0.2177\\
mp&=-131.654+[ND_1 (G)]0.7933
\end{align*}

2. $ND_{2}$ index:\\
\begin{align*}
bp&=-95.9303+[ND_2 (G)]84.2834\\
ct&=47.6081+[ND_2 (G)]98.1177\\
cp&=43.8159-[ND_2 (G)]7.0529\\
mv&=50.7515+[ND_2 (G)]45.1187\\
mr&=6.7917+[ND_2 (G)]13.1956\\
hv&=3.9821+[ND_2 (G)]14.0794\\
st&=9.6530+[ND_2 (G)]4.5432\\
mp&=-150.6221+[ND_2 (G)]17.6077
\end{align*}

3. $ND_{3}$ index:\\
\begin{align*}
bp&=60.7738+[ND_3 (G)]0.0372\\
ct&=220.3722+[ND_3 (G)]0.0508\\
cp&=29.0689-[ND_3 (G)]0.0019\\
mv&=140.8691+[ND_3 (G)]0.0159\\
mr&=32.2248+[ND_3 (G)]0.0054\\
hv&=32.9914+[ND_3 (G)]0.0043\\
st&=18.2408+[ND_3 (G)]0.0020\\
mp&=-118.9098+[ND_3 (G)]0.0078
\end{align*}

4. $ND_{4}$ index:\\
\begin{align*}
bp&=-69.6148+[ND_4 (G)]100.5729\\
ct&=86.3449+[ND_4 (G)]112.5233\\
cp&=42.1443-[ND_4 (G)]8.7145\\
mv&=75.6867+[ND_4 (G)]48.0647\\
mr&=14.6003+[ND_4 (G)]13.7704\\
hv&=9.3610+[ND_4 (G)]16.3338\\
st&=12.1161+[ND_4 (G)]4.8742\\
mp&=-146.0101+[ND_4 (G)]21.4463
\end{align*}

5. $ND_{5}$ index:\\
\begin{align*}
bp&=-62.0831+[ND_5 (G)]11.0891\\
ct&=73.9689+[ND_5 (G)]13.7538\\
cp&=38.1279-[ND_5 (G)]0.7430\\
mv&=74.7263+[ND_5 (G)]5.5594\\
mr&=12.0902+[ND_5 (G)]1.7348\\
hv&=11.8876+[ND_5 (G)]1.7079\\
st&=10.2382+[ND_5 (G)]0.6757\\
mp&=-142.0736+[ND_5 (G)]2.2172
\end{align*}

6. $ND_{6}$ index:\\
\begin{align*}
bp&=35.3191+[ND_6 (G)]0.5065\\
ct&=187.4769+[ND_6 (G)]0.6782\\
cp&=30.8960-[ND_6 (G)]0.0291\\
mv&=127.7664+[ND_6 (G)]0.2304\\
mr&=28.1203+[ND_6 (G)]0.0754\\
hv&=29.6437+[ND_6 (G)]0.0609\\
st&=16.9413+[ND_6 (G)]0.0266\\
mp&=-123.8778+[ND_6 (G)]0.1045
\end{align*}
\newpage
\vfill
Now we use the statistical parameters same as previous discussion to interpret the above regression models, where $N$ denotes the total number of alkanes. 
\begin{table}[ht]
\caption{Statical parameters for the linear QSPR model for $ND_{1}$ (G).}
\centering
\begin{tabular}{|c| c| c| c| c| c| c| c|}
\hline

Physical properties&	N&	c&	m&	r&	SE&	F&	SF\\
\hline
bp&	67&	-8.8069&	4.0114&	0.8160&	22.8878&	129.5828&	4.01E-17\\
\hline

ct&	67&	135.4475&	5.1317&	0.8981&	20.2494&	270.9371&	7.17E-25\\
\hline

cp&	67&	34.6560&	-0.2721&	-0.6941&	2.2736&	60.4267&	7.37E-11\\
\hline

mv&	65&	100.8619&	2.0398&	0.8233&	10.1969&	132.5481&
	3.86E-17\\
	\hline

mr&	65&	20.1480&	0.6398&	0.8683&	2.6501&	193.0535&
	7.53E-21\\
	\hline

hv&	65&	21.8703&	0.5616&	0.7436&	3.6618&
	77.9195&
	1.29E-12\\
	\hline

st&	64&	14.3557&	0.2177&	0.7782&	1.2448&
	95.1879&
	3.83E-14\\
	\hline

mp&	52&	-131.654&	0.7933&	0.2516&	26.4042&
	3.3806&
	0.07191\\

\hline
\end{tabular}
\label{table:11}
\end{table}

\begin{table}[ht]
\caption{Statical parameters for the linear QSPR model for $ND_{2}$(G).}
\centering
\begin{tabular}{|c| c| c| c| c| c| c| c|}
\hline

Physical properties&	N&	c&	m&	r&	SE&	F&	SF\\
\hline
bp&	67&	-95.9303&	84.2834&	0.9020&	22.8878&	129.5828&
	4.01E-17\\
\hline
ct&	67&	47.6081&	98.1177&	0.9033&	20.2494&
	270.9371&
	7.17E-25\\
\hline
cp&	67&	43.8159&	-7.0529&	-0.9465&	2.2737&
	60.4267&
	7.37E-11\\
\hline
mv&	65&	50.7515&	45.1187&	0.9415&	10.1969&
	132.5481&
	3.86E-17\\
\hline
mr&	65&	6.7917&	13.1956&	0.9259&	2.6501&
	193.0535&
	7.53E-21\\
\hline
hv&	65&	3.9821&	14.0794&	0.9638&	3.6618&
	77.9195&
	1.29E-12\\
\hline
st&	64&	9.6530&	4.5432&	0.8090&	1.1652&
	117.4079&
	6.13E-16\\
\hline
mp&	52&	-150.622&	17.6077&	0.2862&	26.1414&
	4.4595&
	0.039726\\

\hline
\end{tabular}
\label{table:12}
\end{table}

\begin{table}[ht]
\caption{Statical parameters for the linear QSPR model for $ND_{3}$(G).}
\centering
\begin{tabular}{|c| c| c| c| c| c| c| c|}
\hline

Physical properties&	N&	c&	m&	r&	SE&	F&	SF\\
\hline
bp&	67&	60.7738&	0.0372&	0.6238&	30.9524&
	41.3958&
	1.71E-08\\
\hline
ct&	67&	220.3722&	0.0508&	0.7318&	31.3747&
	74.9341&
	2E-12\\
\hline
cp&	67&	29.0689&	-0.0019&	-0.3904&	2.9077&
	11.6902&
	0.00109\\
\hline
mv&	65&	140.8691&	0.0159&	0.5684&	14.7809&
	30.0655&
	7.86E-07\\
\hline
mr&	65&	32.2248&	0.0054&	0.6418&	4.0973&
	44.1193&
	8.37E-09\\
\hline
hv&	65&	32.9914&	0.0043&	0.5041&	4.7298&
	21.4635&
	1.86E-05\\
\hline
st&	64&	18.2408&	0.0020&	0.6448&	1.5151&
	44.1117&
	8.93E-09\\
\hline
mp&	52&	-118.9098&	0.0078&	0.2036&	26.7107&
	2.1626&
	0.147672\\
\hline
\end{tabular}
\label{table:13}
\end{table}

\begin{table}[ht]
\caption{Statical parameters for the linear QSPR model for $ND_{4}$(G).}
\centering
\begin{tabular}{|c| c| c| c| c| c| c| c|}
\hline

Physical properties&	N&	c&	m&	r&	SE&	F&	SF\\
\hline
bp&	67&	-69.6148&	100.5729&	0.7947&	24.0365&
	111.4291&
	9.91E-16\\
\hline
ct&	67&	86.3449&	112.5233&	0.7649&	29.6545&
	91.6399&
	4.91E-14\\
\hline
cp&	67&	42.1443&	-8.7145&	-0.8634&	1.5934&
	190.369&
	5.52E-21\\
\hline
mv&	65&	75.6867&	48.0647&	0.7785&	11.2749&
	96.942&
	2.29E-14\\
\hline
mr&	65&	14.6003&	13.7704&	0.7500&	3.5339&
	80.9956&
	6.48E-13\\
\hline
hv&	65&	9.3610&	16.3338&	0.8679&	2.7207&
	192.2597&
	8.3E-21\\
\hline
st&	64&	12.1161&	4.8742&	0.6760&	1.4605&
	52.1868&
	8.8E-10\\
\hline
mp&	52&	-146.0101&	21.4463&	0.2525&	26.3985&
	3.4038&
	0.070974\\

\hline
\end{tabular}
\label{table:14}
\end{table}

\begin{table}[ht]
\caption{Statical parameters for the linear QSPR model for $ND_{5}$(G).}
\centering
\begin{tabular}{|c| c| c| c| c| c| c| c|}
\hline

Physical properties&	N&	c&	m&	r&	SE&	F&	SF\\
\hline
bp&	67&	-62.0831&	11.0891&	0.9166&	15.8369&
	341.4188&
	1.44E-27\\
\hline
ct&	67&	73.9689&	13.7538&	0.9779&	9.6226&
	1422.6559&
	6.55E-46\\
\hline
cp&	67&	38.1279&	-0.7430&	-0.7700&	2.0151&
	94.6835&
	2.61E-14\\
\hline
mv&	65&	74.7263&	5.5594&	0.8881&	8.2565&
	235.2645&
	6.02E-23\\
\hline
mr&	65&	12.0902&	1.7348&	0.9319&	1.9378&
	415.896&
	1.91E-29\\
\hline
hv&	65&	11.8876&	1.7079&	0.8950&	2.443&	253.5998&
	9.12E-24\\
\hline
st&	64&	10.2382&	0.6757&	0.9267&	0.7446&
	377.2765&
	4.73E-28\\
\hline
mp&	52&	-142.0736&	2.2172&	0.2790&	26.4042&
	3.3806&
	0.07191\\
\hline
\end{tabular}
\label{table:15}
\end{table}

\begin{table}[ht]
\caption{Statical parameters for the linear QSPR model for $ND_{6}$(G).}
\centering
\begin{tabular}{|c| c| c| c| c| c| c| c|}
\hline

Physical properties&	N&	c&	m&	r&	SE&	F&	SF\\
\hline
bp&	67&	35.3191&	0.5065&	0.6791&	29.0691&	
	55.6291&
	2.67E-10\\
\hline

ct&	67&	187.4769&	0.6782&	0.7823&	28.6755&
	102.5177&
	5.42E-15\\
\hline

cp&	67&	30.8960&	-0.0291&	-0.4892&	2.7546&
	20.4495&
	2.66E-05\\
\hline

mv&	65&	127.7664&	0.2304&	0.6393&	13.8148&
	43.5369&
	9.97E-09\\
\hline

mr&	65&	28.1203&	0.0754&	0.7033&	3.7979&
	61.6774&
	6.44E-11\\
\hline

hv&	65&	29.6437&	0.0609&	0.5551&	4.5554&
	28.0552&
	1.6E-06\\
\hline

st&	64&	16.9413&	0.0266&	0.6646&	1.4811&
	49.0453&
	2.12E-09\\
\hline

mp&	52&	-123.8778&	0.1045&	0.2192&	26.619&	2.5225&
	0.118536\\

\hline
\end{tabular}
\label{table:16}
\end{table}

Several interesting observations on the data presented in Table \ref{table:3}-\ref{table:16} can be made.
From Table \ref{table:3}, the correlation coefficient of $ND_1$ index with entropy, acentric factor and DHVAP for octane isomers are found to be good (Figure \ref{fig1}). Specially, it is strongly correlated with acentric factor having correlation coefficient $r=-0.9904$. Also, the correlation of this index is good for the physical properties of 67 alkanes except for cp and mp having correlation coefficient values -0.6941 and 0.2516, respectively. The range of correlation coefficient values lies from 0.7436 to 0.8981.

The QSPR analysis of $ND_2$ index reveals that this index is suitable to predict entropy, acentric factor and DHVAP of octane isomers (Figure \ref{fig2}). Also, one can say from Table \ref{table:12} that, this index have remarkably good correlations with the physical properties of alkanes except mp. The correlation coefficients lies from 0.809 to 0.9638 except mp ($r=0.2862$). Surprisingly, the correlation of $ND_2$ with hv is very high with correlation coefficient value 0.9638.\\  
Table \ref{table:13} shows that $ND_3$ index is inadequate for any structure property correlation in case of alkanes having the correlation coefficient values from 0.2036 to 0.7318. But, from Table \ref{table:5}, we can see that $ND_3$ is well correlated with entropy and acentric factor with correlation coefficients -0.9387 and -0.9765 respectively.\\
The QSPR analysis of $ND_4$ index shows that this index is well correlated with entropy, acentric factor, DHVAP, and HVAP for octane isomers (Table \ref{table:6}). Table \ref{table:14} shows that $ND_4$ index is inadequate for structure property correlation in case of alkanes except cp and hv having correlation coefficients -0.8634 and 0.8679, respectively.\\
From Table \ref{table:7}, one can say that $ND_5$ does not sound so good except CP having correlation coefficient 0.8478. But this index can be considered as an useful tool to predict the physical properties of alkanes except cp and mp. This index is suitable to model bp,ct,mv,mr,hv,st with correlation coefficients 0.9166, 0.9779, 0.8881, 0.9319, 0.8950, and 0.9267, respectively.\\
The QSPR analysis of $ND_6$ index reveals that the correlation coefficient of this index with the physical properties of alkanes are very poor (Table \ref{table:8}). The range of correlation coefficient values lies from 0.2192 to 0.7823. But, when we look into the Table \ref{table:16}, we can say that this index has ability to model entropy, acentric factor, and DHVAP for octane isomers.

\section{Correlation with some well-known indices}
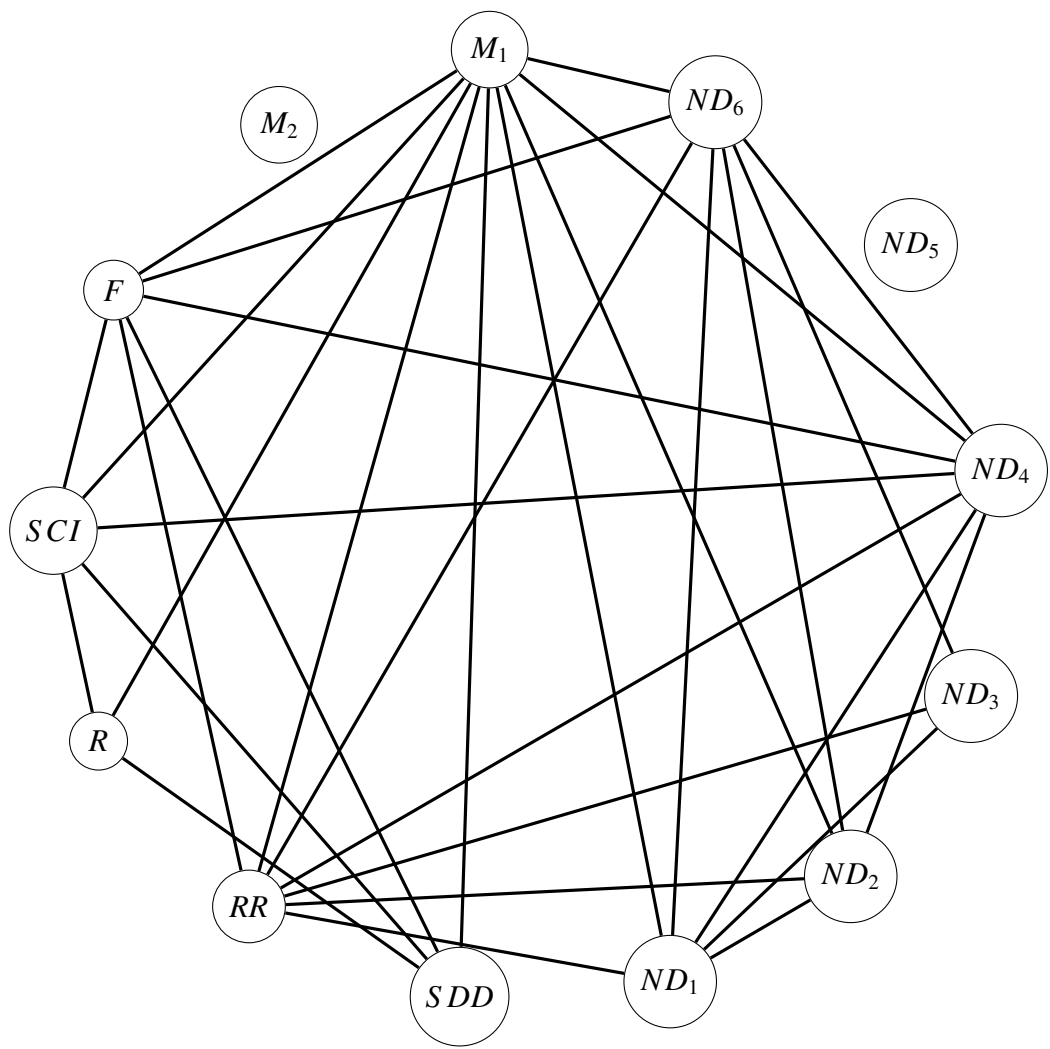
\begin{figure*}[ht]
\begin{center}
\begin{tikzpicture}[scale=.2]
\tikzstyle{every node}=[draw, shape=circle, fill=white, scale=1]
\node (n1) at (50,9) {$ND_1$};
  \node (n2) at (62,16)  {$ND_2$};
  \node (n3) at (70,28) {$ND_3$};
  \node (n6) at (72,43)  {$ND_4$};
  \node (n5) at (66,58) {$ND_5$};
  \node (n4) at (53,67.5)  {$ND_6$};
  \node (n7) at (38,71) {$M_1$};
  \node (n8) at (24,66)  {$M_2$};
  \node (n9) at (13,55) {$F$};
  \node (n10) at (9,39)  {$SCI$};
  \node (n11) at (12,25) {$R$};
  \node (n12) at (22,14)  {$RR$};
  \node (n13) at (36,8) {$SDD$};
\foreach \from/\to in {n1/n2,n1/n3,n1/n4,n1/n6,n1/n7,n1/n12,n2/n4,n2/n6,n2/n7,n2/n12,n3/n4,n3/n12,n4/n6,n4/n7,n4/n9,n4/n12,n6/n7,n6/n9,n6/n10,n6/n12,n7/n9,n10/n11,n10/n13,n10/n7,n10/n9,n11/n7,n11/n13,n12/n7,n12/n9,n13/n7,n13/n9}
\draw [very thick](\from) -- (\to);
\end{tikzpicture}
\caption{Correlation of novel indices with some well-known indices for decane isomer}
\label{fig7}
\end{center}

\end{figure*}

In this section, we investigate the correlation between the new indices and some well-known indices for octane isomers. It is clear from Table \ref{table:17}, that the new indices have high correlation with the well-established indices except $ND_5$ index. Highest correlation coefficient ($r=0.9977$) is between $ND_1$ and $M_{2}$. From table \ref{table:18}, one can say that $ND_5$ has significantly low correlation coefficient with other indices. So we can conclude that $ND_5$ is independent among five indices. A correlation graph (Figure \ref{fig7}) is drawn considering indices as vertices and two vertices are adjacent if and only if $\vert r \vert \geq 0.95$.

\begin{table}[ht]
\caption{Correlation with some well-known indices.}
\centering
\begin{tabular}{|c| c| c| c| c| c| c| c|}
\hline

	&$M_1$&	$M_2$&	$F$&	$SCI$&	$R$&	$RR$&	$SDD$\\
	\hline
$ND_1$&	-0.86958&	-0.8295&	0.995406&	0.817435&	0.946941&	0.741367&	0.937806\\
\hline
$ND_2$&	0.892937&	0.857255&	-0.99282&	-0.83834&	-0.95286&	-0.71895&	-0.93683\\
\hline
$ND_3$&	-0.75331&	-0.70235&	0.953983&	0.691599&	0.865993&	0.746587&	0.861531\\
\hline
$ND_4$&	0.955051&	0.933136&	-0.97643&	-0.9137&	-0.97717&	-0.6806&	-0.95874\\
\hline
$ND_5$&	-0.29763&	-0.22396&	0.665973&	0.198425&	0.471285&	0.56989&	0.461035\\
\hline
$ND_6$&	-0.89201&	-0.85372&	0.994895&	0.848793&	0.964237&	0.739059&	0.959608\\

\hline
\end{tabular}
\label{table:17}
\end{table}

\begin{table}[ht]
\caption{Correlation among new indices.}
\centering
\begin{tabular}{|c| c| c| c| c| c| c| c|}
\hline

	&$ND_1$&	$ND_2$&	$ND_3$&	$ND_4$&	$ND_5$&	$ND_6$\\
	\hline
$ND_1$&	1& & & & & \\ 
\hline					
$ND_2$&	-0.98906&	1& & & & \\	
\hline			
$ND_3$&	0.976497&	-0.94568&	1&&&\\
\hline			
$ND_4$&	-0.95818&	0.980133&	-0.87983&	1&&\\
\hline		
$ND_5$&	0.719609&	-0.68604&	0.825766&	-0.54476&	1&\\
\hline	
$ND_6$&	0.992599&	-0.97954&	0.962921&	-0.95487&	0.675576&	1\\

\hline
\end{tabular}
\label{table:18}
\end{table}

\section{Degeneracy}
The objective of a topological index is to encipher the structural property as much as possible. Different structural formulae should be distinguished by a good topological descriptor. A major drawback of most topological indices is their degeneracy, i.e., two or more isomers possess the same topological index.
\vfill
\clearpage
\FloatBarrier 
Topological indices having high discriminating power captures more structural information. We use the measure of degeneracy known as sensitivity introduced by Konstantinova \cite{19kon96}, which is defined as follows:  
\begin{eqnarray*}
S_{I}=\frac{N-N_{I}}{N},
\end{eqnarray*}

where $N$ is the total number of isomers considered and $N_I$  is the number of them that cannot be distinguished by the topological index $I$. As  $S_I$ increases, the isomer-discrimination power of topological indices increases. The vertex degree based topological indices have more discriminating power in comparison with other classes of molecular descriptors.  For octane and decane isomers, the newly introduced indices exhibit good response among other investigated degree based indices (Table \ref{table:19}).

\begin{table}
  \centering
  \caption{Measure of sensitivity (S_I) of different indices for octane and decane isomers.}
  \begin{tabular}{|c|*{4}{c|}}\hline
    \multirow{2}{*}{Topological indices} & \multicolumn{2}{c|}{ Sensityvity($S_{I}$)} \\\cline{2-3}
                       & Octane & Decane \\ \hline
    $ND_1$&	\textbf{1.000}&	\textbf{1.000}\\
    \hline
$ND_2$&	\textbf{1.000}&	\textbf{1.000}\\
\hline
$ND_3$&	\textbf{1.000}&	\textbf{0.96}\\
\hline
$ND_4$&	\textbf{1.000}&	\textbf{0.987}\\
\hline
$ND_5$&	\textbf{1.000}&	\textbf{0.92}\\
\hline
$ND_6$&	\textbf{0.833}&	\textbf{0.613}\\
\hline
$R$&	0.889&	0.667\\
\hline
$RR$&	0.889&	0.653\\
\hline
$SCI$&	0.889&	0.64\\
\hline
$SDD$&	0.889&	0.547\\
\hline
$M_2$&	0.722&	0.28\\
\hline
$F$&	0.389&	0.133\\
\hline
$M_1$&	0.333&	0.107\\
\hline

  \end{tabular}
  \label{table:19}
  \end{table}

\section{Mathematical properties}

In this section, we discuss about some bounds of the newly proposed indices with some well-known indices. Throughout this section, we consider simple connected graph.  We construct this section with some standard inequalities. We start with the following inequality.  
\begin{lemma}
\label{lem1}
(Radon’s inequality) If $x_{i}, y_{i} > 0, i=1,2,...,n, t>0$, then
\begin{eqnarray}
\label{1}
\frac{\sum\limits_{i = 1}^{n}x_{i}^{t+1}}{\sum\limits_{i = 1}^{n}y_{i}^{t}} \geq \frac{(\sum\limits_{i = 1}^{n}x_{i})^{t+1}}{(\sum\limits_{i = 1}^{n}y_{i})^{t}},
\end{eqnarray}
where equality holds iff $x_i=ky_i$ for some constant $k$,$\forall i=1,2,...,n.$
\end{lemma}

\begin{proposition}
\label{pro1}
For a graph G having $m$ edges with neighbourhood version of second Zagreb index $M_2^{\ast} (G)$ \cite{14sou18}, we have
\begin{eqnarray}
\label{2}
ND_{1}(G) \leq \sqrt{mM_2^{\ast} (G)},
\end{eqnarray}
where equality holds iff $G$ is regular or complete bipartite graph.
\end{proposition}

\n\textit{Proof.} For a graph $G$, $M_2^{\ast} (G)= \sum\limits_{uv \in E(G)}\delta_{G}(u)\delta_{G}(v)$. Now considering $x_{i} =1, y_{i} =\delta_{G}(u)\delta_{G}(v), t = \frac{1}{2}$, in (\ref{1}), we obtain
  \begin{eqnarray}
\label{3}
\frac{\sum\limits_{uv \in E(G)}1}{\sum\limits_{uv \in E(G)}(\delta_{G}(u)\delta_{G}(v))^{\frac{1}{2}}} \geq \frac{(\sum\limits_{uv \in E(G)}1)^{\frac{3}{2}}}{(\sum\limits_{uv \in E(G)}\delta_{G}(u)\delta_{G}(v))^{\frac{1}{2}}}.
\end{eqnarray}

Now using the definition of $ND_{1}$ and $M_2^{\ast}$ indices, we can easily obtain the required bound (\ref{2}). Equality in (\ref{3}) holds iff $\delta_{G}(u)\delta_{G}(v)=k$, a constant $\forall uv \in E(G)$. So the equality in (\ref{2}) holds iff 
$G$ is regular or complete bipartite graph.    \qed  
\FloatBarrier
\begin{lemma}
\label{lem2}
Let $\vec{x}=(x_1,x_2,…,x_n)$ and $\vec{y}=(y_1,y_2,…,y_n)$ be sequence of real numbers. Also let $\vec{z}=(z_1,z_2,…,z_n)$ and $\vec{w}=(w_1,w_2,…,w_n)$ be non-negative sequences. Then 
\begin{eqnarray}
\label{4}
\sum\limits_{i = 1}^{n}w_{i}\sum\limits_{i = 1}^{n}Z_{i}x_{i}^{2}+\sum\limits_{i = 1}^{n}z_{i}\sum\limits_{i = 1}^{n}w_{i}y_{i}^{2} \geq 2 \sum\limits_{i = 1}^{n}z_{i}x_{i}\sum\limits_{i = 1}^{n}w_{i}y_{i},
\end{eqnarray}
In particular, if $z_i$ and $w_i$ are positive, then the equality holds iff $\vec{x}=\vec{y}=\vec{k}$, where $\vec{k}=(k,k,…,k)$, a constant sequence.
\end{lemma}
\begin{proposition}
\label{pro2}
For a graph G having $m$ edges with neighbourhood version of second Zagreb index $M_2^{\ast}(G)$, we have
\begin{eqnarray}
\label{5}
ND_{1}(G) \leq \frac{(m+M_2^{\ast}(G))}{2},
\end{eqnarray}
where equality holds iff G is $P_2$.
\end{proposition}
\n\textit{Proof.}
Considering $x_{i} =\delta_{G}(u)\delta_{G}(v), y_{i} =1, z_{i} =1, w_{i} =1$, in (\ref{4}), we get
  \begin{eqnarray*}
\sum\limits_{uv \in E(G)}1\sum\limits_{uv \in E(G)}\delta_{G}(u)\delta_{G}(v) + \sum\limits_{uv \in E(G)}1\sum\limits_{uv \in E(G)}1 \geq 2 \sum\limits_{uv \in E(G)}\sqrt{\delta_{G}(u)\delta_{G}(v)}\sum\limits_{uv \in E(G)}1.
\end{eqnarray*}
After using the definition of $ND_1$ and $M_2^{\ast}$ indices we can obtain 
\begin{eqnarray*}
m M_2^{\ast}(G)+m^{2} \geq 2mND_1(G).
\end{eqnarray*}
After simplification, the required bound is obvious. \\
From lemma \ref{lem2}, the equality in (\ref{5}) holds iff $\delta_{G}(u)\delta_{G}(v)=1 \forall uv \in E(G)$,i.e. $G$ is $P_2$.    \qed

\textbf{Remark:} By arithmetic mean $\ge$ geometric mean, we can write 
\begin{eqnarray*}
\frac{(m+M_2^{\ast}(G))}{2} \geq \sqrt{m M_{2}^{\ast}(G)}. 
\end{eqnarray*}
So the upper bound of $ND_1 (G)$ obtained in proposition \ref{pro1}, is better than that obtained in proposition \ref{pro2}.

\begin{proposition}
\label{pro3}
For a graph $G$ having second Zagreb index $M_2 (G)$, forgotten topological index $F(G)$ , neighbourhood version of hyper Zagreb index $HM_N (G)$ \cite{14sou18}, neighbourhood Zagreb index $M_N (G)$ \cite{13sou18} , we have 

\begin{eqnarray}
\label{6}
ND_{6}(G) \leq \frac{F(G)}{2} + M_{2}(G)+\frac{HM_{N}(G)}{2}-M_{N}(G),
\end{eqnarray}
equality holds iff $G$ is $P_{2}$.

\end{proposition}

\n\textit{Proof.} For a graph $G$, we have $M_{N}(G) =\sum\limits_{v \in V(G)}\delta_{G}(v)^{2} =\sum\limits_{uv \in E(G)}[\delta_{G}(u)\d_{G}(v)+\delta_{G}(v)\d_{G}(u)]$, $HM_{N}(G) = \sum\limits_{uv \in E(G)}[\delta_{G}(u)+\delta_{G}(v)]^{2}$. We know that for any two non-negative numbers $x, y$,  arithmetic mean $\geq$ geometric mean, i.e., $\frac{x+y}{2} \geq \sqrt{xy},$ equality holds iff $x=y$. Now considering $x=d_G (u)+d_G (v)$, $y=\delta_G (u)+\delta_G (v)$, we get
\begin{eqnarray*}
\frac{[d_G (u)+d_G (v)+\delta_G (u)+\delta_G (v)]}{2} \geq \sqrt{(d_G (u)+d_G (v))(\delta_G (u)+\delta_G (v))},
\end{eqnarray*}
squiring both sides, we have 
\begin{eqnarray*}
4(d_G (u)+d_G (v))(\delta_G (u)+\delta_G (v)) \leq [d_G (u)+d_G (v)+\delta_G (u)+\delta_G (v)]^{2},
\end{eqnarray*} 
which gives
\begin{eqnarray*}
&2\sum\limits_{uv \in E(G)}[(d_{G}(u)\delta_{G}(u) + d_{G}(v)\delta_{G}(v))(d_{G}(u)\delta_{G}(v) + d_{G}(v)\delta_{G}(u))]&\\ &\leq \sum\limits_{uv \in E(G)}[d_{G}(u)^{2}+d_{G}(v)^{2}]+2\sum\limits_{uv \in E(G)}d_{G}(u)d_{G}(v)&\\
& + \sum\limits_{uv \in E(G)}[\delta_{G}(u)^{2}+\delta_{G}(v)^{2}]+2\sum\limits_{uv \in E(G)}\delta_{G}(u)\delta_{G}(v).& 
\end{eqnarray*} 
After simplifying and using the formulation of $ND_6$, $F$, $M_2$, $HM_N$, and $M_N$ indices, the required bound is clear. The equality in (\ref{6}) occurs iff $d_G (u)+d_G (v)= \delta_G (u)+\delta_G (v)$, i.e., $G$ is $P_2$. Hence the proof.   \qed \\\\ 

For a graph $G$ consider 
\begin{align*}
\Delta_{N}&= max \lbrace \delta_{G}(v) : v \in V(G) \rbrace,\\
 \delta_{N}&= min \lbrace \delta_{G}(v) : v \in V(G) \rbrace.
\end{align*}
Thus $\delta_{N} \leq \delta_{G}(u) \leq \Delta_{N}$ for all $u \in V(G)$. Equality holds iff $G$ is regular or complete bipartite graph. Clearly we have the following proposition.
\begin{proposition}
\label{pro4}
For a graph $G$ with $m$ number of edges, we have the following bounds.
\begin{enumerate}
\item[(i)]
$m\delta_{N} \leq ND_{1}(G) \leq m\Delta_{N}$,
\item[(ii)]
$\frac{m}{\sqrt{2\Delta_{N}}} \leq ND_{2}(G) \leq \frac{m}{\sqrt{2\delta_{N}}}$,
\item[(iii)]
$2m\delta_{N}^{3} \leq ND_{3}(G) \leq 2m\Delta_{N}^{3}$,
\item[(iv)]
$\frac{m}{\Delta_{N}} \leq ND_{4}(G) \leq \frac{m}{\delta_{N}}$,
\item[(v)]
$\frac{F_{N}^{\ast}(G)-2M_{2}^{\ast}(G)}{\delta_{N}^{2}}+2m \leq\frac{F_{N}^{\ast}(G)-2M_{2}^{\ast}(G)}{\Delta_{N}^{2}}+2m$,
\end{enumerate}
where \cite{14sou18} $F_{N}^{\ast}(G)= \sum\limits_{uv \in E(G)}[d_{G}(u)^{2}+d_{G}(v)^{2}].$

Equality holds in each case iff $G$ is regular or complete bipartite graph.
\end{proposition}
\begin{lemma}
\label{lem3}
Let $a_i$ and $b_i$ be two sequences of real numbers with $a_i \neq 0$ ($i=1,2,...,n$) and such that $pa_i \leq b_i \leq Pa_i$. Then
\begin{eqnarray}
\label{11}
\sum\limits_{i=1}^{n}b_{i}^{2}+pP\sum\limits_{i=1}^{n}a_{i}^{2} \leq (P+p)\sum\limits_{i=1}^{n}a_{i}b_{i}. 
\end{eqnarray}
 Equality holds iff either $b_i=pa_i$ or $b_i=Pa_i$ for every $i=1,2,...,n.$ 
\end{lemma}
\begin{proposition}
\label{pro5}
For a graph G with m edges having neighbourhood version of second Zagreb index $M_2^{\ast}(G)$, we have 
\begin{eqnarray*}
\label{12}
ND_{1}(G) \geq \frac{M_{2}^{\ast}(G)+m\delta_{N}\Delta_{N}}{\delta_{N}+\Delta_{N}}
\end{eqnarray*}
Equality holds iff $G$ is regular or complete bipartite graph.
\end{proposition}

 \n\textit{Proof.} Putting $a_{i}=1$, $b_{i}=\sqrt{\delta_{G}(u)\delta_{G}(v)}$, $p=\delta_{N}$, $P=\Delta_{N}$ in \ref{11}, we get
 \begin{eqnarray*}
 \sum\limits_{uv \in E(G)}\delta_{G}(u)\delta_{G}(v)+\delta_{N}\Delta_{N}\sum\limits_{uv \in E(G)}1 \leq (\delta_{N}+\Delta_{N})\sum\limits_{uv \in E(G)}\sqrt{\delta_{G}(u)\delta_{G}(v)}.
 \end{eqnarray*}
Now applying the definition of $M_2^{\ast}(G)$, $ND_1(G)$ in the above inequation, we obtain
\begin{eqnarray*}
M_{2}^{\ast}(G) + m\delta_{N}\Delta_{N} \leq (\delta_{N}+\Delta_{N}) ND_{1}(G).
\end{eqnarray*}
Which implies
\begin{eqnarray*}
ND_{1}(G) \geq \frac{M_2^{\ast}(G)+m\delta_{N}\Delta_{N}}{\delta_{N}+\Delta_{N}}.
\end{eqnarray*}
Equality holds iff $\sqrt{\delta_{G}(u)\delta_{G}(v)} = \delta_{N}$ or $\sqrt{\delta_{G}(u)\delta_{G}(v)} = \Delta_{N}$ for all $uv \in E(G)$, i.e. $G$ is regular or complete bipartite graph. Hence the proof.  \qed

\begin{proposition}
For a graph $G$ of size $m$ with fifth version of geometric arithmetic index $GA_{5}$, and second Zagreb index $M_{2}(G)$, we have
\label{ga}
\begin{enumerate}
\item[(i)]
$ND_{5}(G) \geq \frac{2m^{2}}{ GA_{5}}$,
\item[(ii)]
$ND_{5}(G) \geq \frac{4M_{2}(G)^{2}}{m\Delta_{N}^{2}}-2m$.
\end{enumerate}
Equality in both cases hold iff $G$ is regular or complete bipartite graph. 

\end{proposition}
\n\textit{Proof.}
\begin{enumerate}
\item[(i)]
 For a graph $G$, we know that \cite{gra11} $GA_{5}(G)=\sum\limits_{uv \in E(G)}\frac{2\sqrt{\delta_{G}(u)\delta_{G}(v)}}{\delta_{G}(u)+\delta_{G}(v)}$.
Now by Cauchy-schwarz inequality, we have
\begin{eqnarray*}
(\sum\limits_{uv \in E(G)}1)^{2} &=& (\sum\limits_{uv \in E(G)}\sqrt{\frac{\delta_{G}(u)+\delta_{G}(v)}{\sqrt{\delta_{G}(u)\delta_{G}(v)}}} \times \frac{1}{\sqrt{\frac{\delta_{G}(u)+\delta_{G}(v)}{\sqrt{\delta_{G}(u)\delta_{G}(v)}}}})^{2} \\
&\leq& \sum\limits_{uv \in E(G)}\frac{\delta_{G}(u)+\delta_{G}(v)}{\sqrt{\delta_{G}(u)\delta_{G}(v)}}\sum\limits_{uv \in E(G)}\frac{\sqrt{\delta_{G}(u)\delta_{G}(v)}}{\delta_{G}(u)+\delta_{G}(v)}.
\end{eqnarray*}
Thus,
\begin{eqnarray}
\label{e}
2m^{2} \leq GA_{5}(G)\sum\limits_{uv \in E(G)}\frac{\delta_{G}(u)+\delta_{G}(v)}{\sqrt{\delta_{G}(u)\delta_{G}(v)}}.
\end{eqnarray}
We know that
\begin{eqnarray*}
\frac{\delta_{G}(u)}{\delta_{G}(v)}+\frac{\delta_{G}(v)}{\delta_{G}(u)} \geq \sqrt{\frac{\delta_{G}(u)}{\delta_{G}(v)}}+\sqrt{\frac{\delta_{G}(v)}{\delta_{G}(u)}}.
\end{eqnarray*}
From \ref{e}, we obtain $2m^{2} \leq GA_{5}(G)ND_{5}(G)$, i.e. 
\begin{eqnarray*}
ND_{5}(G) \geq \frac{2m^{2}}{ GA_{5}}.
\end{eqnarray*}
Equality holds iff $\frac{\delta_{G}(u)+\delta_{G}(v)}{\sqrt{\delta_{G}(u)\delta_{G}(v)}} = k$, a constant $\forall uv \in E(G)$. That is, $\delta_{G}(u) = some constant \times \delta_{G}(v)$  $\forall uv \in E(G)$, i.e., $G$ is regular or complete bipartite graph. 
\item[(ii)]
By Cauchy schwarz inequality, we have
\begin{eqnarray*}
ND_{5}(G) &=& \sum\limits_{uv \in E(G)}\frac{[\delta_{G}(u)+\delta_{G}(v)]^{2}}{\delta_{G}(u)\delta_{G}(v)}-2m \\
&\geq & \frac{1}{\Delta_{N}^{2}}\sum\limits_{uv \in E(G)}[\delta_{G}(u)+\delta_{G}(v)]^{2}-2m\\
&=&\frac{1}{m\Delta_{N}^{2}}\sum\limits_{uv \in E(G)}1^{2}\sum\limits_{uv \in E(G)}[\delta_{G}(u)+\delta_{G}(v)]^{2}-2m\\
&\geq &\frac{1}{m\Delta_{N}^{2}}[\sum\limits_{uv \in E(G)}(\delta_{G}(u)+\delta_{G}(v))]^{2}-2m = \frac{4M_{2}(G)^{2}}{m\Delta_{N}^{2}}-2m.
\end{eqnarray*}
\end{enumerate}
Equality holds iff $\delta_{G}(u)=\Delta_{N} = \delta_{G}(v)$ and $\delta_{G}(u)+\delta_{G}(v) = c, $ a constant occur simultaneously for all $uv \in E(G).$ That is, $G$ is regular or complete bipartite graph. \\
Hence the proof                \qed

It is obvious that, $\delta_G (u)\geq d_G (u)$ and $\delta_G (v) \geq d_G (v)$, $\forall uv \in E(G)$. Equality appears for $P_2$ only. Keeping in mind this fact, we have the following proposition. 
\begin{proposition}
  For a graph $G$, having Randic index $R(G)$, second Zagreb index $M_2 (G)$, reciprocal Randic index $RR(G)$, sum-connectivity index $SCI(G)$, we have
\begin{enumerate}
\item[(i)]
$ND_{1}(G) \geq RR(G)$
\item[(ii)]
$ND_{2}(G) \geq SCI(G)$
\item[(ii)]
$ND_{3}(G) \geq ReZG_{3}(G)$
\item[(iii)]
$ND_{4}(G) \geq R(G)$
\item[(iv)]
$ND_{5}(G) \leq 2M_{2}(G)$
\item[(v)]
$ND_{6}(G) \leq 2M_{2}(G)$
\end{enumerate}  
  Equality holds in each case iff $G$ is $P_2$.
  
\end{proposition}
\section{Conclusion}
In this article, we have proposed some novel topological indices based on neighbourhood degree sum of end vertices of edges. Their predictive ability have tested using octane isomers and 67 alkanes. It has been shown that these indices can be considered as useful molecular descriptors in QSPR research. These indices are extension of some well-known degree based topological indices (such as $RR$, $SCI$, $SDD$, $R$ etc.). Sometimes the predictive power of these new indices are superior sometimes they are little bit inferior than the old indices. But the degeneracy test on Table $\ref{table:17}$, assures the supremacy of newly designed indices in comparison to the old indices. We have also correlated these indices with other degree based topological indices. This investigation on Table $\ref{table:17}$, $\ref{table:18}$ concludes that $ND_{5}$ index is independent among all novel indices. This work ends with computing some bounds of these novel indices. For further research, these indices can be computed for various graph operations and some composite graphs and networks.


\end{document}